\documentclass[referee]{aa}

\usepackage{graphicx}
\usepackage{natbib}
\usepackage{latexsym}
\usepackage{amsmath}
\usepackage{amssymb}
\usepackage{amstext}
\usepackage{color}
\usepackage{psfrag}

\def\k{km s$^{-1}$}
\def\ks{km s$^{-1}$~}
\def\d{$^\circ$}
\def\m{$^\prime$}
\def\s{$^{\prime\prime}$}
\def\hh{$^{\mathrm h}$}
\def\mm{$^{\mathrm m}$}
\def\ss{$^{\mathrm s}$}
\def\cm3{cm$^{-3}$}
\def\cm2{cm$^{-2}$}

\def\2{$^{12}$CO}
\def\3{$^{13}$CO}
\def\msol{M$_\odot$}
\def\cchi{$\chi^{2}$}

\begin{document}

\title{The environment of the infrared dust bubble N65: a multiwavelength study}
\author {A. Petriella \inst{1}
\and S. Paron \inst{1}
\and E. Giacani  \inst{1}
}

\institute{Instituto de Astronom\'{\i}a y F\'{\i}sica del Espacio (CONICET-UBA),
             CC 67, Suc. 28, 1428 Buenos Aires, Argentina\\
             \email{apetriella@iafe.uba.ar}
}
\offprints{A. Petriella}

   \date{Received <date>; Accepted <date>}
\abstract{}{We investigate the environment of the infrared dust bubble N65 
and search for evidence of triggered star formation in its surroundings. }
{We performed a multiwavelength study of the region around N65 with data taken from large-scale surveys: 
Two Micron All Sky Survey, GLIMPSE, MIPSGAL, SCUBA, and GRS. We analyzed the distribution of the molecular gas
and dust in the environment of N65
and performed infrared photometry and spectral analysis of point sources to search for young stellar objects
and identify the ionizing star candidates. }
{We found a molecular cloud that appears to be fragmented into smaller clumps along the N65 PDR. This 
indicates that the so-called collect and collapse process may be occurring. Several young stellar
objects are distributed among the molecular clumps. They may represent a second generation of stars
whose formation was triggered by the bubble expanding into the molecular gas.
We identified O-type stars inside N65, which are the most reliable ionizing star candidates. }
{}

\titlerunning{The environment of the infrared dust bubble N65}
\authorrunning{A. Petriella et al.}


\maketitle

\section{Introduction}
\label{intro}

Since the early work of \citet{elme77}, it is known that during the supersonic expansion of an HII region, 
a dense layer of material can be collected between the ionization and the shock fronts. 
This layer can be fragmented into massive condensations that may then collapse to form new massive stars and/or clusters. 
Several observational studies have inferred that this process, known as ``collect and collapse'', triggers massive
star formation (see e.g., \citealt{poma09,zav07}, and references therein).
As \citet{deha05} point out, only the presence of either a dense 
molecular shell surrounding the ionized gas of an HII region or massive fragments regularly 
spaced along the ionization front, can prove that we are dealing with the collect and collapse process. 

Using Galactic Legacy Infrared Mid-Plane Survey Extraordinaire (GLIMPSE)
data, \citet{church06,church07} cataloged almost 600 infrared (IR) dust
bubbles, full or partial rings bordered by a photodissociation region (PDR) and detected mainly at 8 $\mu$m, 
which usually enclose ionized gas and hot dust observed at 24 $\mu$m. Most of these bubbles are HII regions 
and several have morphologies that are indicative of triggered star formation.

On the other hand, \citet{cyga08} also using GLIMPSE data, identified more than 300 extended 4.5 $\mu$m sources,
so-called ``extended green objects (EGOs)'', applying the common coding of the [4.5] band as green in three-color
composite Infrared Array Camera images. According to the authors, an EGO is probably a massive young stellar
object (MYSO) driving outflows. The extended emission in the 4.5 $\mu$m band is produced by 
H$_{2}$ ($\nu =$ 0--0, S(9,10,11)) lines and CO ($\nu = 1-0$) band heads, which are excited by the shock of
the outflows propagating in the interstellar medium (ISM). The majority of EGOs are associated with infrared dark
clouds (IRDCs) and some lie over the border of IR dust bubbles.

From the catalog of \citet{church06}, we select the IR dust bubble N65,  
which harbors the EGO G35.03+0.35 in one of its borders (indicative of active star formation),
to study its environment.

N65 is a complete (i.e., closed ring) IR dust bubble centered  
on $\alpha_{2000} =$ 18\hh 54\mm 02\ss, $\delta_{2000} = +$01\d 59\m 30\s~($l = 35$\fdg$000$, $b = +0$\fdg$332$)
with a radius of about 2\m.6 \citep{church06}. Close to N65, over one of its borders, 
centered on $\alpha_{2000} =$ 18\hh 54\mm 04.2\ss, $\delta_{2000} = 
+$02\d 01\m 33.9\s~($l = 35$\fdg$035$, $b = +0$\fdg$338$) lies the IR source IRAS 18515+0157, where 
several molecular lines tracers of star formation were detected. Maser emission from H$_{2}$O and OH was detected by 
\citet{forster89}, as well as methanol \citep{caswell95}. \citet{bronf96} and \citet{jiji99} also detected 
CS and NH$_{3}$, respectively. 
The detection of formaldehyde at v$_{LSR} \sim 57$ \k~was used by \citet{watson03}
to determine that this star-forming region lies at the kinematic 
distance of 10 kpc, the farthest possible given the distance ambiguity of the first Galactic quadrant. 
On the other hand, close to this region, at $l = 35$\fdg$015$, $b = +0$\fdg$356$ and also across the border
of N65, \citet{anderson09} cataloged the UCHII region G35.02+0.35 with a velocity of $\sim 57.2$ \k,
which they locate at the near distance of $\sim 3.6$ kpc. 

In this work, we present a molecular and IR study of the environment surrounding
the IR dust bubble N65 to explore its surrounding ISM and search for signatures of 
star formation. We describe the data used in Sect. \ref{data}, the results and the 
discussion of them are presented in Sect. \ref{resultdisc} and Sect. \ref{sum} summarizes the
results. 

\section{Data}
\label{data}

We analyzed data extracted from five large-scale surveys: the
Two Micron All Sky Survey (2MASS)\footnote{2MASS is a joint project of
the University of Massachusetts and the Infrared Processing and Analysis Center/California Institute of Technology,
funded by the National Aeronautics and Space Administration and the National Science Foundation.},
Galactic Legacy Infrared Mid-Plane Survey Extraordinaire (GLIMPSE),
MIPSGAL, GRS\footnote{Galactic Ring Survey \citep{jackson06}}, and the SCUBA Legacy Catalogue.
GLIMPSE is a mid-infrared survey of the inner 
Galaxy performed with the {\it Spitzer Space Telescope}. We used the mosaicked images from
GLIMPSE and the GLIMPSE Point-Source Catalog (GPSC) in the {\it Spitzer}-IRAC (3.6, 4.5, 5.8, and 8 $\mu$m).
IRAC has an angular resolution of between 1\farcs5 and 1\farcs9 (see \citealt{fazio04} and \citealt{werner04}).
MIPSGAL is a survey of the same region as GLIMPSE, using the MIPS instrument (24 and 70 $\mu$m) on {\it Spitzer}.
The MIPSGAL resolution is 6\s~at 24 $\mu$m. 
The GRS is being performed by the Boston University and the
Five College Radio Astronomy Observatory (FCRAO). The survey maps the galactic Ring in the \3 J=1--0 line
with an angular and spectral resolution of 46\s~and 0.2 \k, respectively (see \citealt{jackson06}).
The observations were performed in both position-switching and on-the-fly mapping modes, achieving an
angular sampling of 22\s. The SCUBA Legacy Catalogues\footnote{http://www2.cadc-ccda.hia-iha.nrc-cnrc.gc.ca/community/scubalegacy/} are two sets of continuum maps and catalogs using 
data at 450 and 850 $\mu$m obtained with the Submillimetre Common User Bolometer Array (SCUBA) with
angular resolutions of 19\s~and 11\s, respectively \citep{difranc08}.

\section{Results and discussion}
\label{resultdisc}

\begin{figure}[h]
\includegraphics[width=14cm]{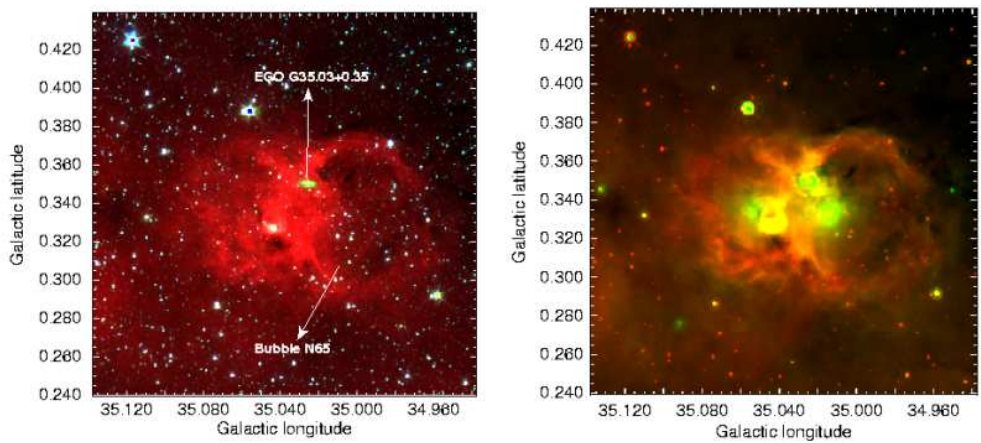}
\caption{Mid-IR emission of the IR dust bubble N65. {\it Left}: {\it Spitzer}-IRAC three-color image 
(3.6 $\mu$m $=$ blue, 4.5 $\mu$m $=$ green, and 8 $\mu$m $=$ red). The location of the EGO G35.03+0.35 is indicated. 
{\it Right}: red is the {\it Spitzer}-IRAC 8 $\mu$m emission and green is the {\it Spitzer}-MIPSGAL emission at 24 $\mu$m.}
\label{iracmips}
\end{figure}

Figure \ref{iracmips} ({\it left}) shows a {\it Spitzer}-IRAC three color image of N65. The three IR bands are 3.6 $\mu$m 
(in blue), 4.5 $\mu$m (in green), and 8 $\mu$m (in red). The location of the EGO G35.03+0.35 
\citep{cyga08} is indicated.
Figure \ref{iracmips} ({\it right}) displays a composite two-color image towards N65. Red and green represent  
the {\it Spitzer}-IRAC emission at 8 $\mu$m and the {\it Spitzer}-MIPSGAL emission at 24 $\mu$m, respectively.
Both figures clearly show the PDR visible in the 8 $\mu$m emission, which originates mainly
in the polycyclic aromatic hydrocarbons (PAHs). The PAH emission delineates the HII region 
boundaries because these large molecules are destroyed inside the ionized region, but are excited in the 
PDR by the radiation leaking from the HII region \citep{poma09}.
The 24 $\mu$m emission, displayed in green in Fig. \ref{iracmips} ({\it right}), corresponds to hot dust, 
distributed mainly towards the eastern border of N65 that coincides with the central position of IRAS 18515+0157.

\begin{figure}[h]
\includegraphics[width=13cm]{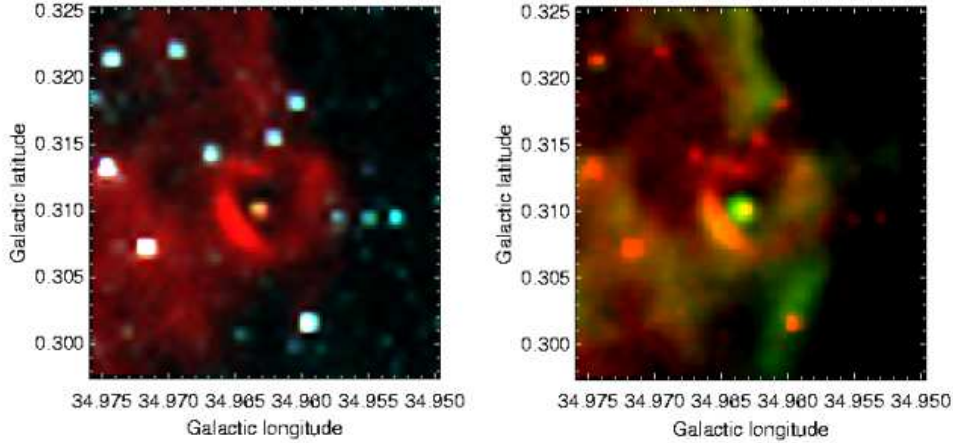}
\caption{Mid-IR emission of the possible smaller IR dust bubble that lies in the southwest border of N65.
{\it Left}: {\it Spitzer}-IRAC three-color image (3.6 $\mu$m $=$ blue, 4.5 $\mu$m $=$ green and 8 $\mu$m $=$ red).
{\it Right}: red is the {\it Spitzer}-IRAC 8 $\mu$m emission and green is the {\it Spitzer}-MIPSGAL
emission at 24 $\mu$m.}
\label{smallbub}
\end{figure}

From Fig. \ref{iracmips}, a probably smaller 
IR dust bubble can be discerned in the southwest border of N65, centered on $l = 34$\fdg$96$, $b = +0$\fdg$31$,
which is displayed in Fig. \ref{smallbub}.
The emission at 8 $\mu$m (Fig. \ref{smallbub} {\it left}) shows an almost circular PDR,
which encloses a region of about 18\s~in diameter. 
At the center of the bubble, a source is visible with emission 
in the 3.5, 4.5, 8, and 24 $\mu$m bands, which is probably the exciting star of this 
structure surrounded by hot dust emitting at 24 $\mu$m
(green in Fig. \ref{smallbub} {\it right}). 

\subsection{Molecular and dust analysis}
\label{molec}

\begin{figure}[h]
\includegraphics[width=13cm]{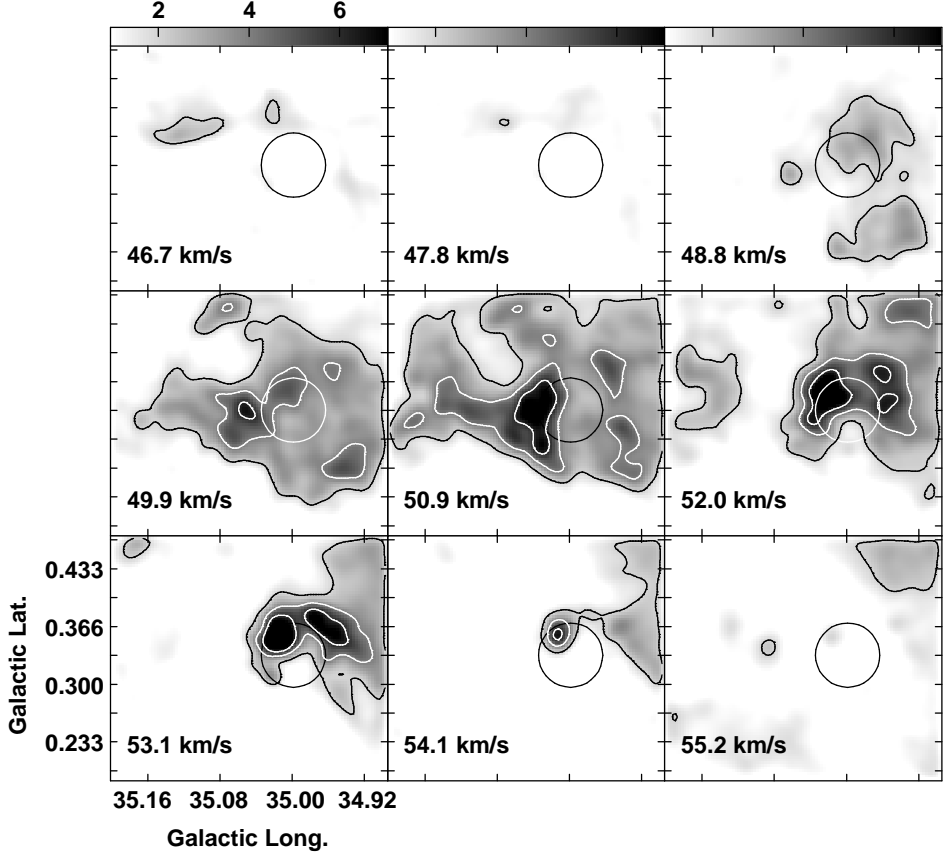}
\caption{Integrated velocity maps of the \3 J=1-0 emission every 1.05 \k. The grey scale is in K \k.
The contour levels are 2, 4 and 6 K \k. The circle represents the position and size of N65.}
\label{molecgas}
\end{figure}

We inspected the molecular gas around N65 from the GRS data in the whole velocity range and found an interesting 
feature around  v $\sim$ 50 \k.
Figure \ref{molecgas} shows the integrated velocity maps of the \3 J=1-0 emission every 1.05 \k~between 
47 and 56 \k. The circle shows the position and size of N65.
A molecular cloud with an arc-like shape open to the south is evident that encloses
the IR bubble over one of its borders. The molecular gas exhibits several clumps along the PDR.
The distribution and morphology of this material suggests that the 
collect and collapse process might be occurring.

\begin{figure}[h]
\centering
\includegraphics[width=9cm]{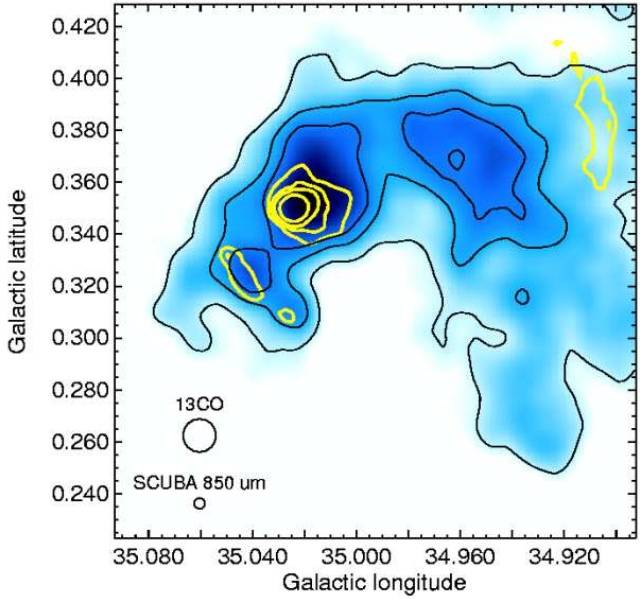}
\caption{\3 emission integrated between 47 and 55 \k, the contours levels are 19, 23, and 33 K \k. The yellow
contours correspond to the 850 $\mu$m continuum emission obtained from SCUBA and the levels are 0.3, 0.6, 0.9, and 2.3
Jy beam$^{-1}$. The beams of each emission are included in the bottom left corner.}
\label{molec+scuba}
\end{figure}

In Fig. \ref{molec+scuba}, we compare the \3 emission integrated within the interval 47 and 55 \k~with 
the continuum dust emission at 850 $\mu$m extracted from the SCUBA (in yellow contours).
In this figure, it is evident that the dust emission coincides with the brightest molecular clump centered
on $l \sim 35\fdg02$, $b \sim 0\fdg35$ and with the position of the EGO G35.03+0.35.
The presence of the EGO together with all the previous results mentioned in Sect. \ref{intro} provide clear 
evidence that this region might be active in star formation.

If we assume that the molecular gas in the velocity range considered before is physically associated 
with N65, we can adopt $\sim 50$ \k~as the systemic velocity of the infrared dust bubble.
According to the galactic rotation model of \citet{fich89} (with $R_{\odot} = 8.5$ kpc and $v_{\odot} = 220$ \k),
we obtain kinematic 
distances of either 3.5 or 10.5 kpc. Considering that the detection rate for IR dust bubbles in
GLIMPSE peaks at the distance of 4.2 kpc within a horizon of 8 kpc \citep{church06}, 
the bubbles at the near kinematic distance are more likely to be detected than those at the far distance. 
Thus, we assume a distance of 3.5 kpc for N65, in coincidence with the UCHII region G35.02+0.35 cataloged
by \citet{anderson09}.

Using the \3 J=1--0 line and assuming local thermodynamic equilibrium (LTE), we estimate the H$_{2}$ column density
toward the brightest molecular clump shown in Fig. \ref{molec+scuba}. We use
$${\rm N(^{13}CO)} = 2.42 \times 10^{14} \frac{T_{\rm ex} \int{\tau_{13} dv}}{1 - exp(-5.29/T_{\rm ex})},$$
to obtain the \3 column density, where $\tau_{13}$ is the optical depth of the line and
following \citet{anderson09} we considered $T_{\rm ex} = 20$ K.
We assume that the \3 emission is optically thin and use the relation N(H$_{2}$)/N(\3)$ \sim 5 \times 10^5$
(e.g., \citealt{simon01}) to estimate a column density of N(H$_{2}$) $\sim 1.6 \times 10^{22}$ cm$^{-2}$.
From this value, we estimate the mass of the molecular clump to be $\sim 2 \times 10^{3}$ \msol.
This value was obtained from
$$ {\rm M} = \mu~m_{{\rm H}} \sum{\left[ D^{2}~\Omega~{\rm N(H_{2})} \right] }, $$
where $\Omega$ is the solid angle subtended by the \3 J=1--0 beam size, $m_{\rm H}$ is the hydrogen mass,
$\mu$ is the mean molecular weight assumed to be 2.8 by taking into account a relative helium abundance
of 25 \%, and $D$ is the distance. Summation was performed over all the observed positions within the
33 K \ks contour level (see Fig. \ref{molec+scuba}). Finally from this mass value, we derive a density of 
$n \sim 10^{4}$ cm$^{-3}$.

Using the 850 $\mu$m continuum emission, we estimate the dust mass from the relation of \citet{tej06}:
$$ M_{dust} = 1.88 \times 10^{-4} \left(\frac{1200}{\nu}\right)^{3+\beta} S_{\nu} \left(e^{0.048 \nu/T_{d}} - 1\right) D^{2},$$
where $S_{\nu}$ is the flux density at the frequency $\nu$. We use 
$S_{353GHz} = 19.42$ Jy as obtained by \citet{difranc08} for this region. We assume that 
$T_{d}$ the dust temperature is 20 K, $\beta$ the dust emissivity index is 2.6 for the assumed
dust temperature in this region according to \citet{hill06}, and $D$ the distance to N65 is 3.5 kpc. 
This equation assumes a standard opacity of $k_{1200GHz} = 0.1$ cm$^{2}$ g$^{-1}$. We obtain a dust mass 
of $M_{dust} \sim 60$ \msol. The dust-to-gas mass ratio is then $\sim 0.03$, which is 3 times higher than the 
canonical Galactic ratio.

\subsection{IR photometry}
\label{photom}
                                                                                                                             
\begin{figure}[h]
\centering
\includegraphics[angle=-90,totalheight=0.35\textheight,viewport=0 0 550 700,clip]{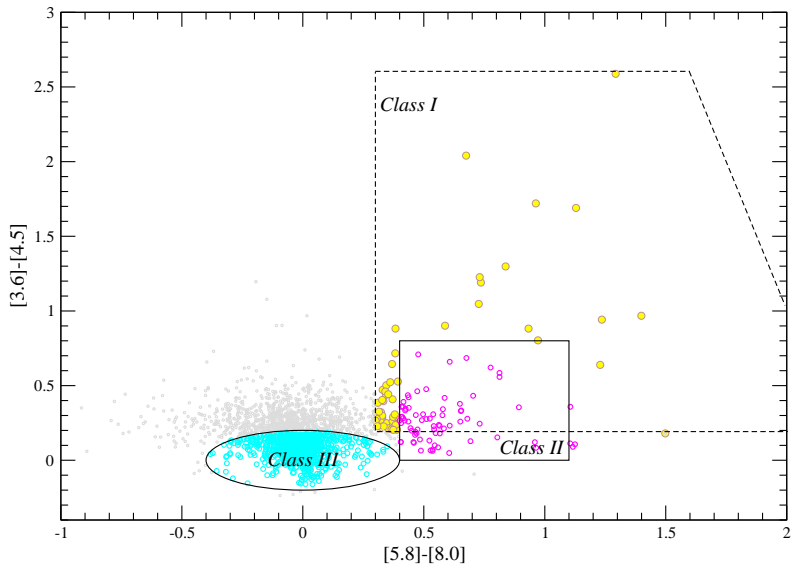}
\caption{\ GLIMPSE color-color diagram [5.8]-[8.0] versus [3.6]-[4.5] for sources within a circle 
of 8\m~in radius centered at N65. We only consider sources with detection in the four {\it Spitzer}-IRAC bands.
The regions indicate the stellar evolutionary stage as defined by \citet{allen04}.
Class I sources are protostar with circumstellar envelopes;
class II are disk dominated objects; and class III are main sequence and giant stars. }
\label{ccdiagr}
\end{figure}

We analyze the distribution of IR point sources in the surroundings of N65 to search 
for signs of star formation and try to identify the ionizing-star candidates.
Figure \ref{ccdiagr} shows the [5.8]-[8.0] versus [3.6]-[4.5]
color-color (CC) diagram for the sources extracted from the GLIMPSE Point Source Catalog in the {\it Spitzer}-IRAC bands
within a circle of 8\m~in radius centered on 
$l = 35$\fdg$00$, $b = +0$\fdg$32$. The size of this region completely covers the extension
of N65 and the surrounding molecular clumps, where stars may be forming. 
We only consider sources with detection in the four bands.
The regions in the figure indicate the stellar evolutionary stages 
based on the criteria of \citet{allen04}: class I sources are protostars with circumstellar envelopes,
class II are disk-dominated objects, and class III are main sequence and giant stars.  
We search for ionizing star candidates among class III sources, while class I and II sources 
are chosen to be the YSO candidates. In the following sections, we report the results of these analyses.  
                                                                                                                             
\subsection{Identifying the exciting star(s) of N65}
\label{ioniz}
                                                                                                                             
We attempt identify or at least propose the possible exciting stars(s) of N65.
                                                                                                                             
\begin{figure}[h]
\centering
\includegraphics[width=8cm,angle=-90]{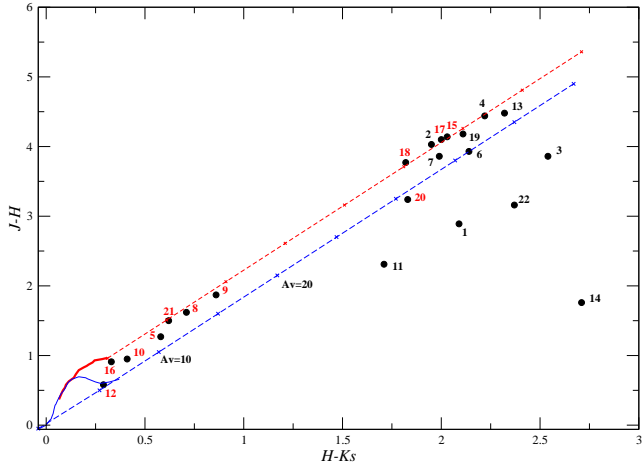}
\caption{({\it H-K}) versus ({\it J-H}) color-color (CC) diagram of the class III objects within the N65 PDR.
The two solid curves represent the location of the main
sequence (thin line) and the giant stars (thicker line) derived from
\citet{bessell88}. The parallel dashed lines are reddening
vectors. We have assumed the interstellar reddening law of \citet{rieke85}
($A_J/A_V$=0.282; $A_H/A_V$=0.175 and $A_K/A_V$=0.112). The sources with red numbers are the ones selected 
to be the candidates for exciting stars of N65.}
\label{cc}
\end{figure}

\begin{figure}[h]
\centering
\includegraphics[width=8cm,angle=-90]{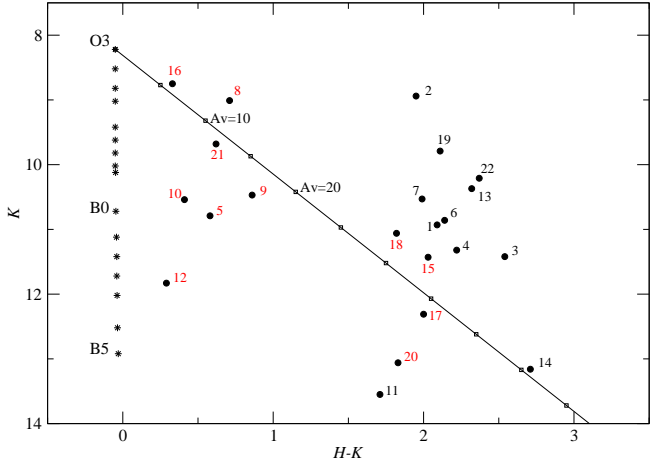}
\caption{({\it H}) versus ({\it H-K}) color-magnitude (CM) diagram of the class III objects within the N65
PDR. The position of the zero-age-main-sequence stars with A$_{v} = 0$ mag at a distance of 3.5 kpc are
indicated by black stars. The reddening curve for an O3 star is shown with a black line. The sources with red
numbers represent are the ones selected to be candidate exciting stars of N65.}
\label{cm}
\end{figure}

\begin{figure}[h]
\centering
\includegraphics[width=8cm]{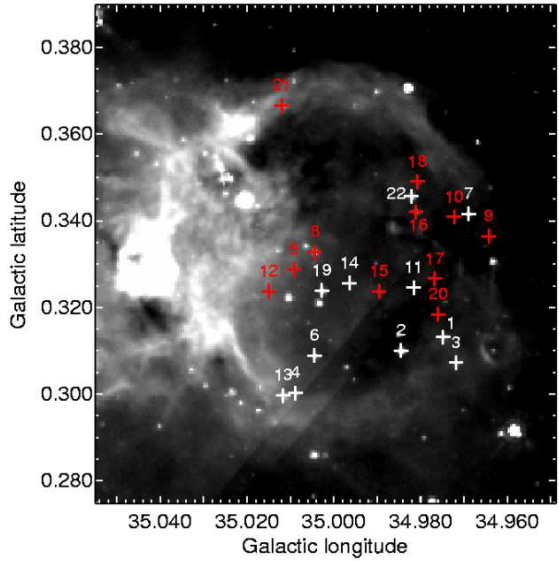}
\caption{Possible exciting star(s) over the 8 $\mu$m emission of N65. The red crosses are the sources selected,
according to the CM and the CC diagrams, to be candidate exciting stars.}
\label{excitstar}
\end{figure}

According to \citet{church06}, IR dust bubbles may be produced by O- and/or B-type stars.
Given that we do not find any cataloged massive star toward N65, we look for the exciting star(s) among
the class III sources because they include main sequence stars.
The ionizing star(s) are not always located close to the geometrical
center of the bubble because of their proper motion. However, we expect to find them inside the
PDR as IR dust bubbles are young objects and the ionizing star(s) may not have enough time
to overcome the shell. To perform near IR photometry using the 2MASS {\it JHK} bands, we select
class III objects within the N65 PDR that have detections in the aforementioned near IR bands.
Figure \ref{cc} shows the ({\it H-K}) versus ({\it J-H}) color-color (CC) diagram of the 22
sources found using this criteria. Figure \ref{cm} shows the {\it K} versus ({\it H-K})
color-magnitude (CM) diagram. Finally, in Fig. \ref{excitstar} we show the location of these sources.
                                                                                                                             
To identify possible exciting star(s) candidates among the 22 sources found,
we select objects that meet both of the following criteria simultaneously: 
(i) according to their position in the CC diagram (Fig. \ref{cc}), the sources
may be main sequence stars and (ii) according to their position in the CM diagram (Fig. \ref{cm}), the sources 
may be O- or B-type stars. These sources are indicated by red numbers in Figs. \ref{cc} and \ref{cm} and
red crosses in Fig. \ref{excitstar}. 

As an additional constraint of the ionizing star(s) candidate, we derive the spectral energy distribution
(SED) of the selected sources by fitting the fluxes in
the IRAC and 2MASS bands with a Kurucz photospheric model using the tool developed by \citet{robit07}
and available online\footnote{http://caravan.astro.wisc.edu/protostars/}.
The fitting tool requires a distance range and a range of visual extinction.
We assume a distance of 3.5 kpc and derive the extinction for each source from the 2MASS
CC diagram {\it (H-K)} versus {\it (J-H)} presented in Fig. \ref{cc}.
Once the SED is obtained, we check whether the fitted effective
temperature is consistent with the temperature of an O- and B-type star.
Our analysis is only qualitative because of the uncertainty in the determination of 
the spectral type from the CM diagram and because the fitting photosphere model extends only to 50,000 K. 

The fitted effective temperatures of stars 5, 10, and  
16 obtained from the SED is $\sim$ 6,000 K and for star 20 is $\sim$ 20,000 K.
These values are too low for a O- or early B-type star. For the remaining sources 8, 9, 12, 15, 17, 18, 20, and 21
the fitted temperature is $\sim$ 50,000 K, which is in closer agreement with the effective temperature 
expected for an O-type star (see \citealt{scha97}). Thus, based on their position inside the bubble, 
we suggest that sources 8, 12, and 15 are the most reliable ionizing-star candidates of N65. 
                                                                                                                                
\subsection{Star formation around N65}
\label{yso}

\begin{figure}[h]
\centering
\includegraphics[width=12cm]{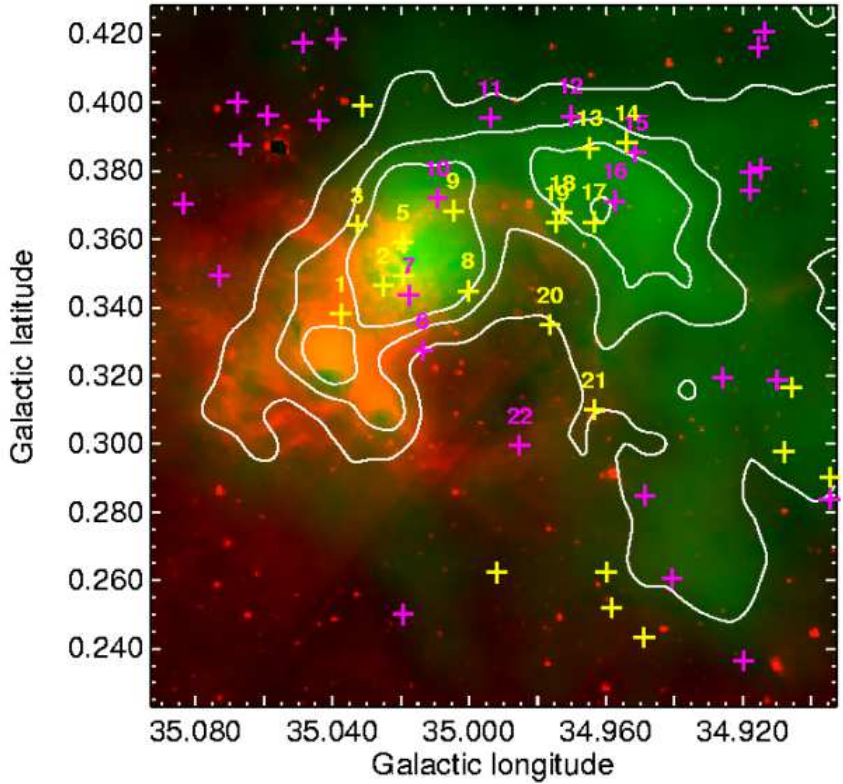}
\caption{\ Two-color image towards N65: \3 emission integrated between 47 and 55 \k (in green with white contours) and
8 $\mu$m emission (in red). We show the distribution of class I (yellow crosses) and class II (magenta crosses)
sources around N65 derived from the CC diagram of Fig. \ref{ccdiagr}. We have labeled the sources located 
upon the molecular cloud and close to the PDR. }
\label{pointsrc}
\end{figure}

Figure \ref{pointsrc} shows the distribution of both class I (yellow crosses) and class II (magenta crosses) point
sources around N65. We note the presence of a group of these sources right upon the
molecular clumps that surround the bubble above its PDR. As molecular clouds are star birth places, we expect
some of these sources to be YSOs whose formation could have been triggered by the expanding
IR dust bubble.
                                                                                                                             
We perform a fitting of the fluxes of YSO candidates in the IRAC and 2MASS bands,
to derive their spectral energy distribution (SED) and constrain
their evolutionary stage. We limit our study to the sources superimposed on the molecular cloud around N65
(see Fig. \ref{pointsrc}). The SED was obtained using the tool developed by
\citet{robit07} available online\footnote{http://caravan.astro.wisc.edu/protostars/}.
This tool consists of a grid of precomputed radiative transfer models with a fast \cchi
~minimization algorithm. We select the models that satisfied the condition
\begin{equation}
\chi^{2} - \chi^{2}_{min} < 2N,
\label{chi2}
\end{equation}
where $\chi^{2}_{min}$~is the minimum value of the \cchi~among all models,
and {\it N} is the number of input data fluxes (fluxes specified as upper limit do not contribute
to {\it N}). Hereafter, we refer to models satisfying Eq. (\ref{chi2}) as ``selected models".
                                                                                                                             
To relate the SED to the evolutionary stage of the YSO, \citet{robit06}
defined three different stages as a complementary classification of Allen's classes. The stage
classification is based on the values of the central source mass $M_\star$, the disk mass $M_{disk}$,
the envelope mass $M_{env}$, and the envelope accretion rate $\dot{M}_{env}$ of the YSO.
Stage I YSOs are those that have $\dot{M}_{env}/M_\star>10^{-6}~yr^{-1}$, i.e., protostars with 
large accretion envelopes;
stage II are those with $\dot{M}_{disk}/M_\star>10^{-6}~yr^{-1}$ and $\dot{M}_{env}/M_\star<10^{-6}~yr^{-1}$, i.e., 
young objects with prominent disks; and stage III are those with $\dot{M}_{disk}/M_\star<10^{-6}~yr^{-1}$ and
$\dot{M}_{env}/M_\star<10^{-6}~yr^{-1}$, i.e., evolved sources where the flux is dominated by the central source.
The stage classification is based on the physical properties of the YSO,
rather than on the spectral index derived from the slope of
its SED. However, as pointed by \citet{robit06}, the stage classification is equivalent to
the class classification for a great number of models. Therefore, an easy way of
identifying young objects is looking for sources that can be fitted only by selected models with $\dot{M}_{env}>0$.
Sources that are fitted by models with zero and non-zero values of $\dot{M}_{env}$ may be
more evolved YSOs whose evolutionary stage is not constrained by the data (see \citealt{poulton08}).
                                                                                                                             
In Table \ref{avelines}, we report the main results of the fitting output for class I and II sources
projected onto the molecular cloud. The YSO identification (Col. 1)
is taken from Fig. \ref{pointsrc}. In Col. 2 and 3, we report the \cchi~per data point of the best-fit model and
the number on models satisfying Eq. (\ref{chi2}), respectively. The remaining columns
report the physical parameters of the source, specifying the range of values of the selected models:
central source mass, disk mass, envelope mass, and envelope accretion rate, respectively.
These results permit us to confirm the presence of young objects in the vicinity of the IR dust bubble N65.

\begin{table}[h]
\caption{Parameters of class I and II YSO candidates derived from the SED fitting.}
\label{avelines}
\centering
\begin{tabular}{ccccccc}
\hline\hline
Source& $\chi^{2}_{min}/{\it N}$  &{\it n} &M$_{\star}$ &M$_{disk}$ &M$_{env}$ &$\dot{M}_{env}$  \\
           &                            &           & (M$_{\odot}$) & (M$_{\odot}$) & (M$_{\odot}$) & (M$_{\odot}/yr$)  \\
\hline
1 & 0.13& 484& $1 - 11$  & $8.3\times{10^{-8}} - 0.4$  & $6.4\times{10^{-9}} - 1800$ & $0 - 0.003$\\
2 & 2.2& 41 & $5 - 8$  & $7.1\times{10^{-4}} - 0.7$  & $1.1\times{10^{-8}} - 1200$ & $0 -  0.002$\\
3 & 0.1& 84 & $1 - 11$& $1.1\times{10^{-6}} - 0.71$   & $1.7\times{10^{-8}} - 140$   & $0 - 0.001$\\
4 & 0.2 & 273 & $0.2 - 11$ & $5.5\times{10^{-5}} - 0.6$ & $3.8\times{10^{-9}} - 360$ & $0 - 0.003$\\
5 & 5 & 4   &     9    & 0                           & 28                   & $3.8\times{10^{-4}}$ \\
6 & 2.2 & 2   &  4.6   & 0.014                       & 37                  &   $2.1\times{10^{-4}}$ \\
7 & 10 & 9 &  $4 - 5$  & $7.4\times{10^{-5}} - 0.02$  & $0.1 - 4$      & $5\times{10^{-6}} - 9\times{10^{-5}} $\\
8 & 1.3 & 1  &   9.6     & 0.75                        & 130                        & $2.1\times{10^{-4}}$\\
9 & 5  & 61 & $2 - 7$& $1.6\times{10^{-7}} - 0.13$    & $1.8\times{10^{-7}} - 17$  & $0 - 5.4\times{10^{-4}}$\\
10& 4  & 9  &  $5$ & $4.5\times{10^{-5}} - 0.001 $ & $0.09 - 1.5$  & $9.5\times{10^{-7}} - 1.4\times{10^{-5}}$ \\
11& 3.3 & 10  &  $2 - 4$   & $6.5\times{10^{-4}} - 0.03$  & $0.02 - 16$  & $4\times{10^{-6}} - 5\times{10^{-4}}$ \\
12& 11 & 2  & $3 -4$    & $0.016 - 0.02$ & $2 - 8 $ &  $ 3.5\times{10^{-6}} - 2\times{10^{-5}}$ \\
13& 2.4 & 13  &  $1 - 7$ & $1.1\times{10^{-6}} - 0.3 $ & $0.006 - 43$   & $3.5\times{10^{-8}} - 1.1\times{10^{-4}}$ \\
14& 1.4 & 298 & $1 - 11$ & $3\times{10^{-6}} - 0.7$ & $5\times{10^{-9}} - 140$  & $ 0 - 0.002$\\
15& 1   &  3  &  $8 - 9$ & $0.03 - 0.2 $             & $70 - 140$         &  $1.4\times{10^{-8}} - 2\times{10^{-4}} $ \\
16& 0.03 & 430 &  $1 - 8$ & $3\times{10^{-8}} - 0.4 $ & $1\times{10^{-8}} - 23$  & $0 - 4\times{10^{-4}}$\\
17& 3   & 2   & 4.6      & 0.014                       & 37             & $2.1\times{10^{-4}}$ \\
18& 0.8 & 13 & $2 - 9$   & $6.7\times{10^{-4}} - 0.7$ & $0.1 - 140$ & $3\times{10^{-6}} - 4\times{10^{-4}} $  \\
19& 0.4 & 85& $2 - 21$   &         $0 - 1.1$         & $0.4 - 1300$       & $4.1\times{10^{-5}} - 0.002$\\
20& 2& 2 & $9 - 10$    & $0.3 - 0.8$ &      $65 - 130$ & $2\times{10^{-4}}$ \\
21&0.003& 1652& $0.7 - 15$  & $6\times{10^{-7}} - 0.9 $ & $4.5\times{10^{-9}} - 560$  & $0 - 0.001$ \\
22& 1.3& 50 & $3 - 15$& $9\times{10^{-9}} - 0.2$ & $1.4\times{10^{-8}} - 200$ & $ 0 - 6\times{10^{-4}}$ \\
\hline
\end{tabular}
\end{table}

We identify 12 sources that are only fitted by models with $\dot{M}_{env}>0$; they are sources 5, 6, 7, 10,
11, 12, 13, 15, 17, 18, 19, and 20. These are presumably young objects embedded in prominent envelopes.
In every case, the selected models are prevalently stage I with a few stage II.
The only exception is source 13, which is fitted by both stage I and stage III models.

Sources 5, 15, and 20 appear to be the youngest YSOs. They may all be massive central sources (between 8 and 10 \msol)
surrounded by envelopes of several solar masses. In the case of 15 and 20, their ages are estimated to be 
$\sim{10^{4}}$ years. Regarding the source 5, its position is coincident with the ultra-compact HII region G35.02+0.35 and
its age is about $10^{3}$ years. This is a really young massive source and a clear sign that star formation is
still occurring in the vicinity of N65.

The selected models (i.e., models satisfying Eq. \ref{chi2}) for sources 7, 10, 11, and 12 spread
over a similar range of values among them. These YSOs contain
a central source of mass less than 5 \msol~and ages between $10^{4}$ and $10^{5}$ years.

For source 19, we obtain a large number of selected models (85) because for this source we only fitted the four
IRAC fluxes. However, all of them are stage I so this source is probably a young object. The large spread in the
range of parameters does not allow us to constraint its mass and age.

Projected onto the molecular cloud, we also identify sources with zero and non-zero values
for $\dot{M}_{env}$: sources 1, 2, 3, 4, 9, 14, 16, 21, and 22. The evolutionary stage of these sources
cannot be assured without doubt because the parameters of the selected models are spread over a wide range.
We should include fluxes measured at other wavelengths to reduce the number of models satisfying
the condition defined in Eq. (\ref{chi2}). However, we can identify sources that are in their early stages
of evolution (embedded or disk dominated) because the selected models are either stage I or II. 
This is the case for sources 1, 2, 4, 14, and 16.
For the remaining sources, 3, 9, 16, and 22, the selected models are distributed between stages I, II, and III so their
evolutionary stage cannot be established without controversy.

We note that source 21 is located at the center of the smaller IR bubble
introduced in Sect. \ref{resultdisc} that lies right upon the PDR of N65 (see Fig. \ref{smallbub}).
We fitted the SED of the source and obtained more than
1600 models satisfying Eq. (\ref{chi2}) and, consequently, a wide range of physical parameters for the YSO,
so we cannot establish its evolutionary stage.
However, its location in the center of the smaller IR dust bubbles suggests that
this source may be a second generation massive star blowing a stellar
wind strong enough to evacuate a bubble around itself. More observations are needed to investigate this scenario.

In conclusion, the SED study confirms the existence of several YSOs possibly embedded in the molecular clumps that
surround N65, which is indicative of the collect and collapse process occurring in this region.

\subsection{SED analysis of the EGO G35.03+0.35}
\label{ego}
                                       
\begin{figure}[h]
\centering
\includegraphics[height=5.5cm]{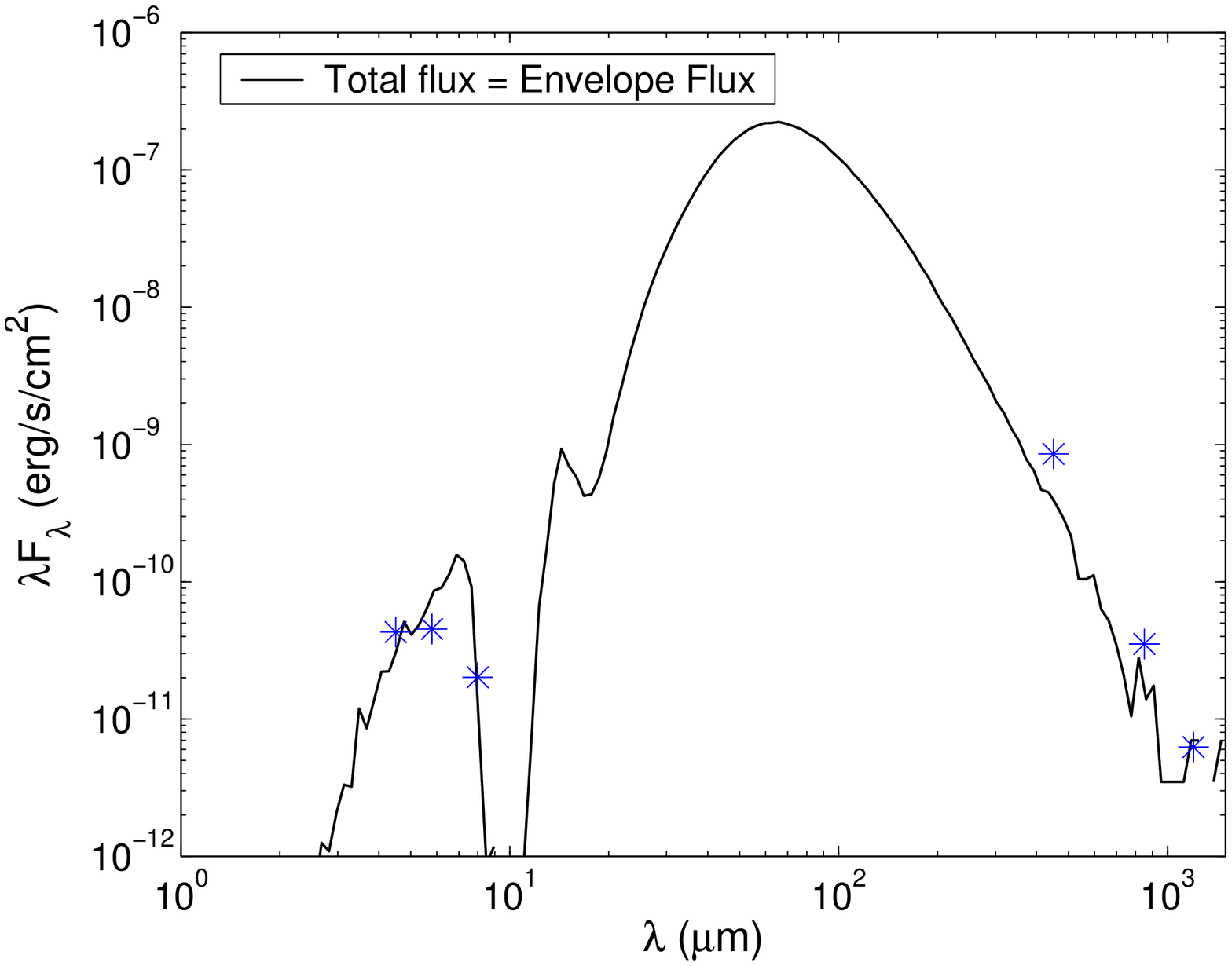}
\includegraphics[height=5.5cm]{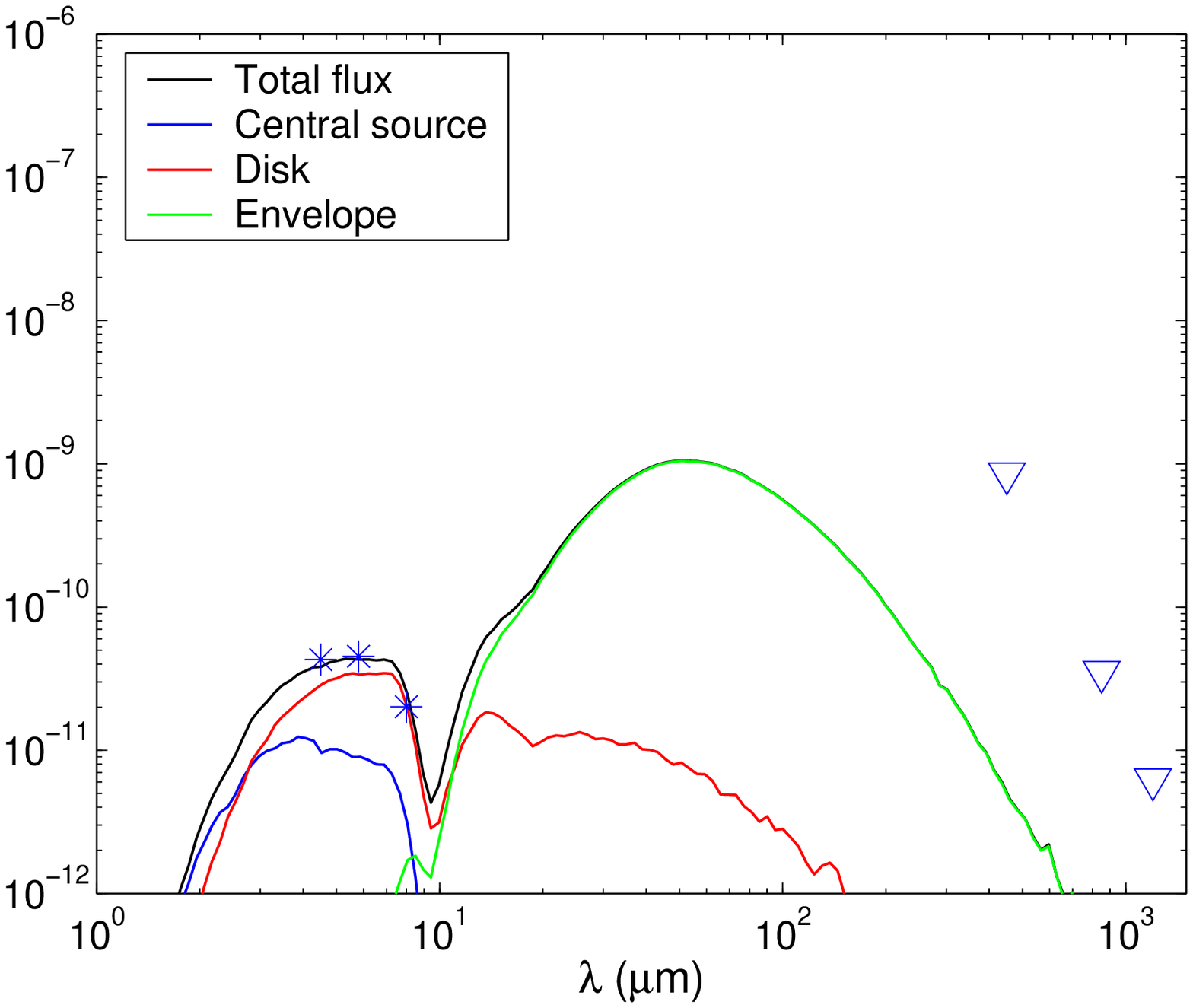}
\caption{\ Best-fit SED model for the EGO G35.05+0.35. The points in blue are the measured fluxes at
the {\it Spitzer}-IRAC bands of 4.5, 5.8, and 8.0 $\mu$m, the SCUBA bands of 450 and 850 $\mu$m, and the band of
1,200 $\mu$m measured with the SIMBA bolometer at SEST.
{\it Left}: SCUBA and SIMBA data are assumed to represent flux emitted only by the EGO.
The total flux (black line) is completely dominated by the
flux from the envelope as would be expected in a young stellar object at its earlier stages of evolution.
{\it Right}: SCUBA and SIMBA data are taken to be an upper limit. 
The total flux at the longer wavelengths is dominated by the
envelope, while at the shorter wavelengths the main contributions originates in the disk (red line)
and the central source (blue line). }
\label{sedEGO}
\end{figure}

We derive the SED for the EGO G35.05+0.35 using the fluxes in the {\it Spitzer}-IRAC 4.5, 5.8, and 8.0 $\mu$m bands
(the GPSC reports no detection in the 3.6 $\mu$m band and the source is absent in the 2MASS catalog). We also use
fluxes in the far IR of the SCUBA bands of 450 and 850 $\mu$m  and in the band of
1200 $\mu$m measured by the SIMBA bolometer at SEST \citep{hill06}. Both the SCUBA and SIMBA datasets
have a lower angular resolution than that of the GPSC and the measured fluxes may include
contributions from other sources near the EGO. For this reason, we perform a fitting of the SED by considering
two different cases: a$)$ assuming that the far IR data represent the flux emitted only by the EGO, the SED
of the best-fit model being shown in Fig. \ref{sedEGO} ({\it left}); and b$)$ using the
far IR data as an upper limit (i.e., assuming that the emission originates in the EGO and, probably,
other sources around it), the SED of the best-fit model in this case being shown in Fig. \ref{sedEGO} ({\it right}).
In Table \ref{tablaEGO}, we report the physical parameters of the selected models,
considering both SCUBA and SIMBA data as flux emitted only by the EGO (case {\it a}) and upper limits (case {\it b}).
The column designation is the same as in Table \ref{avelines}.   

\begin{table}[h]
\caption{Physical parameters of EGO G35.05+0.35 derived from the SED fitting for the selected models satisfying
Eq. (\ref{chi2}). }
\label{tablaEGO}
\centering
\begin{tabular}{ccccccc}
\hline\hline
{\it Case} & $\chi^{2}_{min}/{\it N}$  &{\it n} &M$_{\star}$ &M$_{disk}$ &M$_{env}$ &$\dot{M}_{env}$  \\
     &              &           &  (M$_{\odot}$)  & (M$_{\odot}$) & (M$_{\odot}$) & (M$_{\odot}/yr$)  \\
\hline
a & 30  & 2  & $27 - 33$  & $0 - 0.3$  & $1900 - 2500$ & $ 0.004 -  0.006$\\
b & 3.3 & 16 & $6 - 10$  & $0.005 - 0.8$  & $0.7- 1100$ & $ 2\times{10^{-5}} - 0.002  $\\
\hline
\end{tabular}
\end{table}

According to Fig. \ref{sedEGO} ({\it left}), if we consider fluxes at the longer wavelengths
to be emitted only by the EGO, the total flux of the best-fit model is completely dominated by the envelope. 
As pointed out by \citet{robit06}, the envelope of
sources with high accretion rates becomes optically thick. In our case of $\dot{M}_{env}\sim10^{-3}$~M$_{\odot}/yr$,
radiation emitted
by both the central source and the disk is probably absorbed and re-emitted by the envelope of gas and dust.
As can be seen in Table \ref{tablaEGO}, the physical parameters of the EGO indicate that it is a massive
central source surrounded by a massive envelope. 
According to Fig. \ref{sedEGO} ({\it right}), if we take SCUBA and SIMBA measurements to be upper limits, the total flux 
of the best-fit model is still dominated by the envelope but only at the longer wavelengths. In the near and 
mid IR, the main contribution to the total flux comes from the disk.
The accretion rate values that we derived for the envelope are 
lower than those derived in the previous case (see Table \ref{tablaEGO}).
Thus, the envelope is expected to be more transparent to radiation coming from the inner
regions (disk and central source).

Both cases show that the fitted SED of the EGO is indicative of a massive central source
surrounded by a massive envelope. The age of the YSO is estimated in between $10^{4}$~and $10^{5}$~years. 
From these results, we suggest that the EGO G35.05+0.35 is a candidate massive young stellar
object at the earlier stages of evolution with 
outflowing activity, in agreement with that proposed by \citet{cyga08}.
The presence of this object is a clear sign that star
formation is occurring in the vicinity of N65.
                                                                                                                             
\section{Summary}
\label{sum}

We have conducted a study of the surroundings of the IR dust bubble N65 using large-scale surveys
and archival data in the radio and infrared bands. 
The main results can be summarized as follows:

\begin{enumerate}
\item We discovered the presence of a molecular cloud with an arc-like shape enclosing N65
over its east, north and west borders and centered on a velocity of $\sim 50$ \k.
The cloud appears to be fragmented into smaller molecular clumps that are seen right upon the PDR.
Based on the distribution and morphology of the molecular gas between 47 and 55 \k, we
suggest that the collect and collapse process may be occurring.
\item The EGO G35.05+0.35 lies in the most intense molecular clump, which belongs to the
discovered molecular shell. From the \3 J=1--0 emission, we estimated its molecular mass and density to be 
$\sim 2 \times 10^{3}$ \msol~and $\sim 10^{4}$ cm$^{-3}$, respectively. We found that the
dust continuum emission, traced by the 850 $\mu$m band, also peaks in this region. We
estimated a dust-to-gas ratio of $\sim 0.003$ for this clump.   
\item From the analysis of the infrared photometry of point sources around N65,
we identified three O- or early B-type stars as the most probable candidates responsible for the creation of N65.
In addition, we found many YSOs distributed right upon
the molecular clumps around N65. The analysis of the SED of these sources confirms that many of them
are in the embedded or disk-dominated stage. This suggests that the gas collected into the molecular
clumps may have collapsed to form a second generation of stars, showing another galactic case
of the collect and collapse process.
\item The SED of EGO G35.05+0.35 as derived by fitting the fluxes at the mid and far IR, confirms
that it is a massive young stellar object.
\end{enumerate}

\section*{Acknowledgments}
                            
A.P. is a doctoral fellow of CONICET, Argentina.                                                         
S.P. and E.G. are members of the {\sl Carrera del investigador cient\'\i fico} of CONICET, Argentina. 
This work was partially supported by the CONICET grant PIP 112-200801-02166, UBACYT A023 and ANPCYT
PICT-2007-00902.

\bibliographystyle{aa}  
\bibliography{biblio}
\IfFileExists{\jobname.bbl}{}
{\typeout{}
\typeout{****************************************************}
\typeout{****************************************************}
\typeout{** Please run "bibtex \jobname" to optain}
\typeout{** the bibliography and then re-run LaTeX}
\typeout{** twice to fix the references!}
\typeout{****************************************************}
\typeout{****************************************************}
\typeout{}
}

\label{lastpage}
 
\end{document}